\documentclass[twocolumn,prl,aps,amsfonts,showpacs]{revtex4}

\usepackage{graphicx}

\begin{document}
\draft
\preprint{}

\newcommand{\1}{{\bf \scriptstyle 1}\!\!{1}}
\newcommand{\I}{{\rm i}}
\newcommand{\p}{\partial}
\newcommand{\D}{^{\dagger}}
\newcommand{\bs}{{\bf s}}
\newcommand{\bx}{{\bf x}}
\newcommand{\br}{{\bf r}}
\newcommand{\bk}{{\bf k}}
\newcommand{\bv}{{\bf v}}
\newcommand{\bp}{{\bf p}}
\newcommand{\bu}{{\bf u}}
\newcommand{\bA}{{\bf A}}
\newcommand{\bB}{{\bf B}}
\newcommand{\bE}{{\bf E}}
\newcommand{\bF}{{\bf F}}
\newcommand{\bI}{{\bf I}}
\newcommand{\bK}{{\bf K}}
\newcommand{\bL}{{\bf L}}
\newcommand{\bP}{{\bf P}}
\newcommand{\bQ}{{\bf Q}}
\newcommand{\bS}{{\bf S}}
\newcommand{\bH}{{\bf H}}
\newcommand{\balpha}{\mbox{\boldmath $\alpha$}}
\newcommand{\bsigma}{\mbox{\boldmath $\sigma$}}
\newcommand{\bSigma}{\mbox{\boldmath $\Sigma$}}
\newcommand{\bOmega}{\mbox{\boldmath $\Omega$}}
\newcommand{\bpi}{\mbox{\boldmath $\pi$}}
\newcommand{\bphi}{\mbox{\boldmath $\phi$}}
\newcommand{\bnabla}{\mbox{\boldmath $\nabla$}}
\newcommand{\bmu}{\mbox{\boldmath $\mu$}}
\newcommand{\bepsilon}{\mbox{\boldmath $\epsilon$}}

\newcommand{\iLambda}{{\it \Lambda}}
\newcommand{\cA}{{\cal A}}
\newcommand{\cD}{{\cal D}}
\newcommand{\cF}{{\cal F}}
\newcommand{\cL}{{\cal L}}
\newcommand{\cH}{{\cal H}}
\newcommand{\cI}{{\cal I}}
\newcommand{\cM}{{\cal M}}
\newcommand{\cO}{{\cal O}}
\newcommand{\cR}{{\cal R}}
\newcommand{\cU}{{\cal U}}
\newcommand{\cT}{{\cal T}}

\newcommand{\be}{\begin{equation}}
\newcommand{\ee}{\end{equation}}
\newcommand{\bea}{\begin{eqnarray}}
\newcommand{\eea}{\end{eqnarray}}
\newcommand{\beqa}{\begin{eqnarray*}}
\newcommand{\eeqa}{\end{eqnarray*}}
\newcommand{\nn}{\nonumber}
\newcommand{\DD}{\displaystyle}

\newcommand{\ba}{\left[\begin{array}{c}}
\newcommand{\baa}{\left[\begin{array}{cc}}
\newcommand{\baaa}{\left[\begin{array}{ccc}}
\newcommand{\baaaa}{\left[\begin{array}{cccc}}
\newcommand{\ea}{\end{array}\right]}

\title{Fault-tolerant quantum computing with spins using the conditional Faraday rotation}

\author{Michael N.~Leuenberger
}
\affiliation{Department of Physics, University of California San Diego, La Jolla, CA 92093}

\date{\today}

\begin{abstract}
We propose a fault-tolerant scheme for deterministic quantum computing with spins that is based on a quantum teleportation scheme using the conditional Faraday rotation.  
The phase gate between two sets of noninteracting quantum dots, embedded in microcavities inside a photonic crystal, is mediated by single photons, which yields a Faraday rotation rate high enough for gate operation times of 100 ps. Using sets of quantum dots and error correction codes makes our scheme fault-tolerant. Single-qubit operations on encoded qubits can be implemented by means of the optical Stark effect combined with the optical RKKY interaction.
\end{abstract}


\maketitle

The idea of using spin in quantum dots for quantum computing\cite{Loss,Ziese,Awschalom} has become very promising since 
time-resolved Faraday and Kerr rotation measurements in GaAs semiconductors revealed an electron spin coherence length and time exceeding 100~$\mu$m and  100~ns, respectively\cite{Kikkawa1998,Kikkawa1999}. In quantum dots spin lifetimes of 20 ms have been reported recently\cite{Kroutvar}. Besides the method of electrically controlled gates, there have been several proposals on reaching optical gate control, such as quantum electrodynamical interaction between quantum dots in a microdisk cavity\cite{Imamoglu} and optical RKKY interaction between quantum dots\cite{Piermarocchi}.
Implementations of optical gate control include optically induced entanglement and phase shifts between two excitons in a quantum dot\cite{Chen,Li}.
Recently, the probabilistic CNOT gate for all-optical quantum computing based on a teleportation protocol\cite{Knill} has been implemented experimentally\cite{OBrien}. An interferometric approach to linear-optical deterministic CNOT gate for photonic qubits has recently been demonstrated experimentally\cite{Fiorentino}.

Here we propose a scheme for deterministic quantum computing with spins in quantum dots using the Faraday rotation of single photons due to the nonresonant interaction of the photons with the electrons in the valence band states. We make use of a recently proposed teleportation scheme using the conditional Faraday rotation\cite{Leuenberger}, where it has been shown that the Greenberger-Horne-Zeilinger state\cite{GHZ} in the spin-photon-spin system can be produced in a quantum system consisting of two spins, each surrounded by a microcavity, and a single photon. We show that this entanglement can be used to induce deterministically a conditional phase gate between the two spin qubits. We then introduce fault-tolerance by performing the quantum computations on two sets of spins. Each set of spins consists of $n$ noninteracting spins of single excess electrons of $n$ quantum dots. The qubit redundancy provides a protection against qubit flip errors. For protection against phase errors, the quantum information must be encoded into the entanglement of several qubits, such as shown by Shor for a nine-qubit code\cite{Shor1995}. We implement Shor's quantum error correction code with $n=9$ spins. For this we use three microcavites that each contain three quantum dots. For the implementation of the single-qubit operations on the nine-qubit code we propose to combine the optical Stark effect\cite{Gupta} with the optical RKKY interaction\cite{Piermarocchi}. We show that the initialization and the read-out of the encoded qubits can be performed by means of the conditional Faraday rotation.

We describe now in detail our quantum computing scheme (see Fig.~\ref{Coupling}). We start with the phase gate for single qubits. Let us define two persons Alice and Bob. Both of them have one photonic crystal, in which $n$ noninteracting quantum dots are embedded.
The single excess electrons of Alice's quantum dots are in a general single-spin state 
$\left|\psi\right>_{\rm A}=\alpha\left|\uparrow\right>_{\rm A}+\beta\left|\downarrow\right>_{\rm A}$, where the quantization axis is the $z$ axis. 
Bob's spins are in a general single-spin state 
$\left|\psi\right>_{\rm B}=\gamma\left|\uparrow\right>_{\rm B}+\delta\left|\downarrow\right>_{\rm B}$.
The photons that interact with both Alice's and Bob's quantum dots are initially in a horizontal linear polarization state
$\left|\leftrightarrow\right>$.

Since there is no need to detect optically the spin states in transverse direction\cite{Leuenberger,Leuenberger2004}, our quantum dots can be non-spherical.
So each photon can virtually create only a heavy-hole exciton on each quantum dot (see Fig.~\ref{Selection_rules}). 
The strong selection rules imply that only a $\sigma_{(z)}^+$ ($\sigma_{(z)}^-$) photon can interact with the quantum dot if the excess electron's spin is up (down). This leads to a conditional Faraday rotation of the linear polarization of the photon depending on the spin state of the quantum dot.
The photonic crystal ensures that the photon's propagation direction is always in $z$ direction, perpendicular to the quantum dot plane.

The interaction Hamiltonian of the photon-quantum dot system in the rotating frame reads
\be
\cH_{\rm ep}=\baaaa 
\hbar\omega_{\rm d} & V_{\rm hh} & 0 & 0 \\
V_{\rm hh} & 0 & 0 & 0 \\
0 & 0 & \hbar\omega_{\rm d} & V_{\rm hh} \\
0 & 0 & V_{\rm hh} & 0
\ea
\ee
with the basis states $\left|\uparrow,{\rm hhx}\right>$, $\left|\uparrow\right>\left|\sigma_{(z)}^+\right>$,
$\left|\downarrow,{\rm hhx}\right>$, $\left|\downarrow\right>\left|\sigma_{(z)}^-\right>$. 
First, the spin on Alice's quantum dot is prepared in the state 
$\left|\psi_{\rm e}(0)\right>_{\rm A}=\alpha\left|\uparrow\right>_{\rm A}+\beta\left|\downarrow\right>_{\rm A}$.
So we start with the electron-photon state $\left|\psi_{\rm ep}(0)\right>=\left(\alpha\left|\uparrow\right>_{\rm A}+\beta\left|\downarrow\right>_{\rm A}\right)\left|\leftrightarrow\right>$.
Exact evaluation of the time evolution $\left|\psi_{\rm ep}(t)\right>=e^{-\frac{i}{\hbar}\cH_{\rm ep}t}\left|\psi_{\rm ep}(0)\right>$ yields
\bea
\left|\psi_{\rm ep}(t)\right> & = & \frac{\alpha}{\sqrt 2}\left\{
\frac{e^{-\frac{i\omega_{\rm d}t}{2}}}{E}\left[-2V_{\rm hh}\sin\left(\frac{Et}{2\hbar}\right)\left|\uparrow,{\rm hhx}\right> \right.\right. \nn\\
& & +\left.\left(\hbar\omega_{\rm d}\sin\left(\frac{Et}{2\hbar}\right)
-E\cos\left(\frac{Et}{2\hbar}\right)\right)\left|\uparrow\right>\left|\sigma_{(z)}^+\right>\right] \nn\\
& & +\left.\left|\uparrow\right>\left|\sigma_{(z)}^-\right>\right\} \nn\\
& & +\frac{\beta}{\sqrt 2}\left\{
\frac{e^{-\frac{i\omega_{\rm d}t}{2}}}{E}\left[-2V_{\rm hh}\sin\left(\frac{Et}{2\hbar}\right)\left|\downarrow,{\rm hhx}\right> \right.\right. \nn\\
& & +\left.\left(\hbar\omega_{\rm d}\sin\left(\frac{Et}{2\hbar}\right)
-E\cos\left(\frac{Et}{2\hbar}\right)\right)\left|\downarrow\right>\left|\sigma_{(z)}^-\right>\right] \nn\\
& & +\left.\left|\downarrow\right>\left|\sigma_{(z)}^+\right>\right\},
\eea
where $E=\sqrt{4V_{\rm hh}^2+(\hbar\omega_{\rm d})^2}$.
Since we do not want to produce an exciton after the interaction time $T$ of the photon with the quantum dot, we require that
$\frac{ET}{2\hbar}=j\pi$, where $j$ is a positive integer.
The uncertainty of the interaction time $T$ must be much smaller than $2\pi\hbar/E=10$ ps, which can be satisfied with a 0.1 ps mirror switch (see below).
The resulting electron-photon state is
\be
\left|\psi_{\rm{Ap}}(T)\right> = \alpha\left|\uparrow\right>_{\rm A}\left|-\frac{\omega_{\rm d}T}{4}\right> 
+\beta\left|\downarrow\right>_{\rm A}\left|+\frac{\omega_{\rm d}T}{4}\right>,
\label{scattered}
\ee
where 
$\left|\varphi\right>=\left({e^{-i\varphi}\left|{\sigma_{(z)}^+}\right>+e^{i\varphi}\left|{\sigma_{(z)}^-}\right>}\right)/\sqrt 2$. 
While in Ref.~\cite{Leuenberger} the conditional Faraday rotation is produced by nonresonant interaction under the condition $V_{\rm hh}\ll\hbar\omega_{\rm d}$, we get here much closer to resonance where the spin-photon interaction satisfying $\frac{ET}{2\hbar}=j\pi$ produces an enhanced conditional Faraday rotation around the $z$ axis by the angle $\pm \omega_{\rm d}T/4$. If $\omega_{\rm d}T/2=(2l+1)\pi/2$, where $l=0,1,2,\cdots$, the linear polarization of the incoming photon is rotated $-(2l+1)\pi/4$ by the spin up component, and at the same time is rotated $+(2l+1)\pi/4$ by the spin down component, yielding two orthogonal polarizations. Thus 
$\left|\psi_{\rm{Ap}}(T)\right>=\left(\alpha\left|\uparrow\right>_{\rm A}\left|\searrow\hspace{-0.35cm}\nwarrow\right> 
+ \beta\left|\downarrow\right>_{\rm A}\left|\nearrow\hspace{-0.35cm}\swarrow\right>\right)/\sqrt 2$
for $l=0,2,4,\cdots$
or $\left|\psi_{\rm{Ap}}(T)\right>=\left(\alpha\left|\uparrow\right>_{\rm A}\left|\nearrow\hspace{-0.35cm}\swarrow\right> 
+ \beta\left|\downarrow\right>_{\rm A}\left|\searrow\hspace{-0.35cm}\nwarrow\right>\right)/\sqrt 2$
for $l=1,3,5,\cdots$, which is maximally entangled.  

\begin{figure}[htb]
\includegraphics[width=8cm]{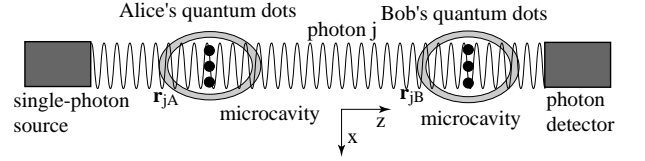}
\caption[]{The nonresonant interaction of the photons with each set of Alice's and Bob's quantum dots produces the conditional phase gate required for universal quantum computing. For Shor's nine-qubit error correction code three of Alice's quantum dots at $\br_{\rm jA}$ are connected to three of Bob's quantum dots at $\br_{\rm jB}$ through the single photon $j$ for $j=1,2,3$.}
\label{Coupling}
\end{figure}

Now we let the photon interact with Bob's quantum dots, which yields
\bea
\left|\psi_{\rm{ApB}}(2T)\right> & = & \left(\alpha\gamma\left|\uparrow_{\rm A}\right>\left|-\frac{\omega_{\rm d}T}{4}-\pi/4\right>\left|\uparrow_{\rm B}\right> \right. \nn\\
& & + \alpha\delta\left|\uparrow_{\rm A}\right>\left|-\frac{\omega_{\rm d}T}{4}+\pi/4\right>\left|\downarrow_{\rm B}\right> \nn\\
& & + \beta\gamma\left|\downarrow_{\rm A}\right>\left|+\frac{\omega_{\rm d}T}{4}-\pi/4\right>\left|\uparrow_{\rm B}\right> \nn\\
& & + \left.\beta\delta\left|\downarrow_{\rm A}\right>\left|+\frac{\omega_{\rm d}T}{4}+\pi/4\right>\left|\downarrow_{\rm B}\right>\right). 
\eea
Choosing $\omega_{\rm d}T=(2l+1)\pi$, we obtain
\bea
\left|\psi_{\rm{ApB}}(2T)\right> & = & \left|\updownarrow\right>
\left(-\alpha\gamma\left|\uparrow_{\rm A}\right>\left|\uparrow_{\rm B}\right> + \beta\delta\left|\downarrow_{\rm A}\right>\left|\downarrow_{\rm B}\right>\right)
\nn\\
& & + \left|\leftrightarrow\right>\left(\alpha\delta\left|\uparrow_{\rm A}\right>\left|\downarrow_{\rm B}\right> + 
\beta\gamma\left|\downarrow_{\rm A}\right>\left|\uparrow_{\rm B}\right>\right). \label{updownrep}
\eea
If the linear polarization of the photon is measured in the $\nearrow\hspace{-0.35cm}\swarrow$ axis, we obtain the outcome of  the conditional quantum phase gate between single spins
\bea
\left|\psi_{\rm{AB}}(2T)\right> & = & 
-\alpha\gamma\left|\uparrow_{\rm A}\right>\left|\uparrow_{\rm B}\right> + \beta\delta\left|\downarrow_{\rm A}\right>\left|\downarrow_{\rm B}\right>
\nn\\
& & +\alpha\delta\left|\uparrow_{\rm A}\right>\left|\downarrow_{\rm B}\right> + 
\beta\gamma\left|\downarrow_{\rm A}\right>\left|\uparrow_{\rm B}\right>, \label{phasegate}
\eea
where the conditional phase shift is $\pi$ if both spins are up.

\begin{figure}[htb]
\includegraphics[width=8cm]{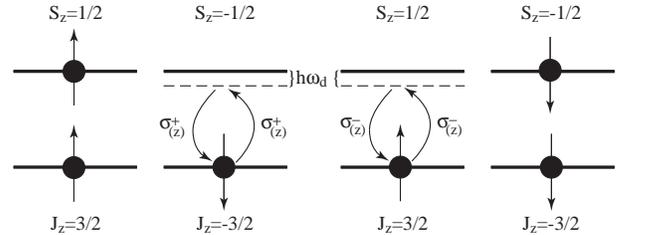}
\caption[]{Selection rules for nonresonant interaction of the photon with the quantum dot. If the spin is up (down), only a $\sigma_{(z)}^+$ ($\sigma_{(z)}^-$) photon can interact with the quantum dot.}
\label{Selection_rules}
\end{figure}

For universal quantum computing the conditional phase gate described above and single-qubit operations are sufficient\cite{Rauschenbeutel,Zubairy,Pazy}. The single-qubit operations on Alice's and Bob's spins can be implemented by means of the optical Stark effect\cite{Gupta,Pryor}.

A very efficient way to protect Alice's and Bob's qubits from decoherence is to use quantum error correcting codes. The simplest code was introduced by Shor\cite{Shor1995}. His idea is to encode a single qubit into nine qubits, with the mapping
$\left|\leftarrow\right>\rightarrow\left|000\right>=\frac{1}{2^{3/2}}
\left(\left|\leftarrow\leftarrow\leftarrow\right>+\left|\rightarrow\rightarrow\rightarrow\right>\right)^{\otimes 3}$,
$\left|\rightarrow\right>\rightarrow\left|111\right>=\frac{1}{2^{3/2}}
\left(\left|\leftarrow\leftarrow\leftarrow\right>-\left|\rightarrow\rightarrow\rightarrow\right>\right)^{\otimes 3}$, where we use the $S_x$ representation $\left|\leftarrow\right>=\left(\left|\uparrow\right>+\left|\downarrow\right>\right)/{\sqrt 2}$ and 
$\left|\rightarrow\right>=\left(\left|\uparrow\right>-\left|\downarrow\right>\right)/{\sqrt 2}$.
In this way, the quantum information is converted from the local single qubit into the three times redundant nonlocal entanglement between three qubits. A requirement for this nine-qubit code to work is that the errors are local and not correlated between qubits, which is satisfied in our quantum computing scheme because all the quantum dots do not interact in the case of the phase gate. The meaning of uncorrelated errors is changed in the case of the single-qubit operations on the nine-qubit code (see below).
Our goal is to find out if the conditional Faraday rotation can also produce a conditional two-qubit phase gate on $\left|000\right>$ and $\left|111\right>$. It will turn out to be possible. 

Each photon connects three of Alice's quantum dots with three of Bob's quantum dots. So we need three photons to perform a conditional phase gate on $\left|000\right>$ and $\left|111\right>$.
We change now to the $S_z$ representation, which gives 
$\left|0\right>
=\frac{1}{2}\left(\left|\uparrow\uparrow\uparrow\right>+\left|\uparrow\downarrow\downarrow\right>
+\left|\downarrow\uparrow\downarrow\right>+\left|\downarrow\downarrow\uparrow\right>\right)$ and
$\left|1\right>
=\frac{1}{2}\left(\left|\downarrow\downarrow\downarrow\right>+\left|\uparrow\uparrow\downarrow\right>
+\left|\uparrow\downarrow\uparrow\right>+\left|\downarrow\uparrow\uparrow\right>\right)$.
We also define the states $\left|\bar 0\right>
=\frac{1}{2}\left(\left|\uparrow\uparrow\uparrow\right>-\left|\uparrow\downarrow\downarrow\right>
-\left|\downarrow\uparrow\downarrow\right>-\left|\downarrow\downarrow\uparrow\right>\right)$ and
$\left|\bar 1\right>
=\frac{1}{2}\left(\left|\downarrow\downarrow\downarrow\right>-\left|\uparrow\uparrow\downarrow\right>
-\left|\uparrow\downarrow\uparrow\right>-\left|\downarrow\uparrow\uparrow\right>\right)$.
The interaction of Alice's three quantum dots with a photon by $\omega_{\rm d}T=(2l+1)\pi$ transforms $\left|\psi_{\rm Ap}(0)\right>=\left(\alpha\left|0\right>_{\rm A}+\beta\left|1\right>_{\rm A}\right)\left|\leftrightarrow\right>$ into the entangled state
\bea
\left|\psi_{\rm Ap}(T)\right> & = & \frac{\alpha}{2}
\left[\left|\uparrow\uparrow\uparrow\right>\left|-3\pi/4\right>_{z}
+\left(\left|\uparrow\downarrow\downarrow\right>+\left|\downarrow\uparrow\downarrow\right>\right.\right. \nn\\
& & +\left.\left.\left|\downarrow\downarrow\uparrow\right>\right)\left|+\pi/4\right>_{z}\right]
+\frac{\beta}{2}
\left[\left(\left|\uparrow\uparrow\downarrow\right>+\left|\uparrow\downarrow\uparrow\right>\right.\right. \nn\\
& & \left.+\left.\left|\downarrow\uparrow\uparrow\right>\right)\left|-\pi/4\right>_{z}
+\left|\downarrow\downarrow\downarrow\right>\left|+3\pi/4\right>_{z}\right] \nn\\
& = & \alpha\left|\bar 0\right>\left|\nearrow\hspace{-0.35cm}\swarrow\right>+\beta\left|\bar 1\right>\left|\searrow\hspace{-0.35cm}\nwarrow\right>
\label{Faraday_odd}
\eea
for $l=0,2,4,\cdots$, or 
\be
\left|\psi_{\rm Ap}(T)\right>=\alpha\left|\bar 0\right>_{\rm A}\left|\searrow\hspace{-0.35cm}\nwarrow\right>+\beta\left|\bar 1\right>_{\rm A}\left|\nearrow\hspace{-0.35cm}\swarrow\right>
\label{Faraday_even}
\ee
for $l=1,3,5,\cdots$.
If we let the photon also interact with Bob's three quantum dots and measure its polarization in the $\nearrow\hspace{-0.35cm}\swarrow$ axis, we obtain
\bea
\left|\psi_{\rm{AB}}(2T)\right> & = & 
-\alpha\gamma\left|\bar 0\right>_{\rm A}\left|\bar 0\right>_{\rm B} + \beta\delta\left|\bar 1\right>_{\rm A}\left|\bar 1\right>_{\rm B}
\nn\\
& & +\alpha\delta\left|\bar 0\right>_{\rm A}\left|\bar 1\right>_{\rm B} + 
\beta\gamma\left|\bar 1\right>_{\rm A}\left|\bar 0\right>_{\rm B}.
\eea
To restore the original states $\left|0\right>$ and $\left|1\right>$, another two $\leftrightarrow$-polarized photons have to pass independently through Alice's and Bob's three quantum dots (with $\omega_{\rm d}T=(2l+1)\pi$) and be measured in the $\leftrightarrow$-polarization, which yields
\bea
\left|\psi_{\rm{AB}}(3T)\right> & = & 
-\alpha\gamma\left|0\right>_{\rm A}\left|0\right>_{\rm B} + \beta\delta\left|1\right>_{\rm A}\left|1\right>_{\rm B}
\nn\\
& & +\alpha\delta\left|0\right>_{\rm A}\left|1\right>_{\rm B} + 
\beta\gamma\left|1\right>_{\rm A}\left|0\right>_{\rm B}. \label{phasegate_codes}
\eea
This procedure can be applied in parallel to the three sets of three quantum dots in $\left|000\right>$ and $\left|111\right>$, giving rise to a $\pi$ phase shift only for $\left|000\right>_{\rm A}\left|000\right>_{\rm B}$, since $e^{3i\pi}=e^{i\pi}=-1$. 
This means it is possible to implement the conditional phase gate using the conditional Faraday rotation also for encoded qubits.

For universal quantum computation it is also important to know how to initialize a qubit state. In particular, we want to produce a cat state of the form $\left|0\right>=\frac{1}{\sqrt 2}\left(\left|\leftarrow\leftarrow\leftarrow\right>+\left|\rightarrow\rightarrow\rightarrow\right>\right)$, which is required for Shor's nine-qubit code. This can be done probabilistically also by means of the conditional Faraday rotation. Let us start from the state 
$\left|\psi_{\rm e}(0)\right>=\left|\leftarrow\leftarrow\leftarrow\right>=(\left|0\right>+\left|1\right>)/\sqrt{2}$, which can be produced by using a $\leftrightarrow$-linearly polarized laser that excites electron-hole pairs and by subsequent tunneling of the hole out of the quantum dot on picosecond timescales\cite{Kroutvar}. Similar to Eqs.~(\ref{Faraday_odd}) and (\ref{Faraday_even}), after the nonresonant interaction of the three quantum dots with a photon traveling in $z$ direction we obtain
$\left|\psi_{\rm ep}(T)\right>=\frac{1}{\sqrt 2}\left(\left|\bar 0\right>\left|\nearrow\hspace{-0.35cm}\swarrow\right>+\left|\bar 1\right>\left|\searrow\hspace{-0.35cm}\nwarrow\right>\right)$
for $l=0,2,4,\cdots$, or $\left|\psi_{\rm ep}(T)\right>=\left(\left|\bar 0\right>\left|\searrow\hspace{-0.35cm}\nwarrow\right>
+\left|\bar 1\right>\left|\nearrow\hspace{-0.35cm}\swarrow\right>\right)/\sqrt 2$ for $l=1,3,5,\cdots$.
For e.g. $l=1,3,5,\cdots$, measuring the polarization of the photon in $\searrow\hspace{-0.35cm}\nwarrow$ direction yields the cat state $\left|\bar 0\right>$ on the three quantum dots with 50\% probability. After the interaction of another $\leftrightarrow$-polarized single photon with the three quantum dots, the spin state is restored to $\left|0\right>$. Applying this method to all the three sets of three quantum dots completes the initialization of the nine-qubit state $\left|\psi_9\right>=\left|000\right>$. Although the probability for successful intitialization is $1/8$, the initialization process can be repeated as many times as necessary.

The single-qubit operations for encoded nine qubits can be performed by means of the optical Stark effect\cite{Gupta,Pryor}. 
The idea is to first produce a strong exchange interaction $\cH_{\rm ex}=J(\bs_1\cdot\bs_2+\bs_2\cdot\bs_3)$ between the three spins by means of the optical RKKY interaction\cite{Piermarocchi}. The ferromagnetic exchange $J$ must be so strong that the three spins act as a single spin of length $S=3/2$. Then the spin-orbit interaction of the form $\cH_{SO}=\lambda \bL\cdot\bS$, with $\lambda\ll J$, leads to the anisotropic spin Hamiltonian $\cH_{\rm S}=DS_x^2$ due to the uniaxial symmetry of the three quantum dots, where $D$ is negative because of the ferromagnetic exchange\cite{Yosida}. $\cH_{\rm S}$ lifts partially the four-fold degeneracy, leaving the states 
$\left|M_S=+ 3/2\right>=\left|\leftarrow\leftarrow\leftarrow\right>$ and 
$\left|M_S=-3/2\right>=\left|\rightarrow\rightarrow\rightarrow\right>$ degenerate. 
The coupling to the environment through e.g. spin-phonon or hyperfine interaction must be much weaker than $D$, in order to avoid decoherence that is stronger than for a spin-1/2 qubit. In other words, the errors are not correlated between the spin-3/2 levels.
Also the states $\left|0\right>$ and $\left|1\right>$ are degenerate. By applying a circularly polarized nonresonant laser beam propagating in $z$ direction, which gives rise to an effective magnetic field $\bB=(0,0,B_z)$ in $z$ direction, phase oscillations between $\left|0\right>$ and $\left|1\right>$ can be induced. Rabi oscillations can be induced by a linearly polarized laser beam propagating in $z$ direction, which produces an effective magnetic field $\bB=(B_x,0,0)$ in $x$ direction. In order to avoid mixing of the states $\left|M_S=\pm 1/2\right>$, we have to ensure that $B_z,B_x\ll D$.

The last requirement for universal quantum computing is the ability to read out a result of the form $\left|\psi_{\rm r}(0)\right>=\alpha_r\left|000\right>+\beta_r\left|111\right>$. Let a photon interact with one of the three sets of three quantum dots for $T=(2l+1)\pi/\omega_{\rm d}$. Then, for e.g. $l=1,3,5,\cdots$, we obtain $\left|\psi_{\rm r}(T)\right>=\alpha_r\left|\bar 000\right>\left|\searrow\hspace{-0.35cm}\nwarrow\right>+\beta_r\left|\bar 111\right>\left|\nearrow\hspace{-0.35cm}\swarrow\right>$. Measuring the polarization of the photon in the $\searrow\hspace{-0.35cm}\nwarrow$ axis leads to a detection with probability $|\alpha_r|^2$ and to no detection with probability $|\beta_r|^2$. This detection method can be applied to the second and third set of three quantum dots separately, as required for fault-tolerant quantum computing. 

We calculate now the numerical values for the experimental implementation of the phase gate.
From the requirement that no exciton is produced, yielding $\frac{ET}{2\hbar}=j\pi$, and the Faraday rotation angle $\omega_{\rm d}T/4=(2l+1)\pi/4$, we obtain a pulse duration of 
\be
T=\frac{(2l+1)\pi}{\omega_{\rm d}}
\ee
and the detuning energy
\be
\hbar\omega_{\rm d}=\frac{2V_{\rm hh}}{\sqrt{\frac{4j^2}{(2l+1)^2-1}-1}}.
\ee
The size of a microcavity within a photonic crystal can be as small as 0.04 $\mu$m$^3$ (see Ref.~\cite{Yoshie}). Assuming that the microcavity is cubical, we get a size of $L=0.35$~$\mu$m. The mirrors of the microcavity can be actively $Q$-switched. Since an all-optical switch responds nowadays with 0.1 ps precision\cite{Friberg,Naruse}, the pulse duration of the photon must also be about $T_p=0.1$~ps, which has e.g. a length of $L_p=10$~$\mu$m in ZnSe and $L_p=8.6$~$\mu$m in GaAs. A microcavity of dimensions $L_x=L_y=1$~$\mu$m and $L_z=10$~$\mu$m gives rise to an electron-photon interaction energy of $V_{\rm hh}=58$~$\mu$eV in ZnSe and $V_{\rm hh}=36$~$\mu$eV in GaAs, given an oscillator strength of $f=166$ (see Ref.~\cite{Guest}). 
In order to reduce the bandwidth $\Gamma_{\rm{photon}}$ of the photon pulse down to $4$~$\mu$eV in ZnSe and $2$~$\mu$eV in GaAs, we choose $j=6$ and $l=5$.
Then we get an interaction time of $T=85$ ps in ZnSe and $T=140$ ps in GaAs and a detuning energy of $\hbar\omega_{\rm d}=0.3$~meV in ZnSe and $\hbar\omega_{\rm d}=0.2$~meV in GaAs, giving rise to a phase error of $\Gamma_{\rm{photon}}/\hbar\omega_{\rm d}=1$~\%.
Due to the photonic bandgap, the decay time of excitons can be as long as $\tau=10$~ns\cite{Lodahl}, which corresponds to a linewidth of $\Gamma=0.03$~$\mu$eV. Therefore the probability for the photon to leak out of the microcavity during the interaction time is $1-e^{-T/\tau}=1$~\%.

In order to perform a single-qubit operation on the nine-qubit code in 100 ps, the magnetic field should be around $H=0.1$ T, corresponding to a Zeeman energy of $B=6$ $\mu$eV. Since $B_z,B_x\ll D\ll J$, the anisotropy energy has to be about $D=0.1$ meV and the exchange coupling must be about $J=1.0$ meV, which can be achieved by the optical RKKY interaction\cite{Piermarocchi}. Since 0.1 meV corresponds to a temperature of 1 K, the single-qubit operation should be performed below about 0.1 K. The larger is the anisotropy energy, the faster the single-qubit operation can be performed.

In conclusion, we have shown that the conditional Faraday rotation produced by the nonresonant interaction of the photon with two quantum dots gives rise to a phase gate between the two qubits represented by the excess spins in the quantum dots. By tuning the photon's frequency very close to the bandgap it is possible to perform the phase gate on a time scale of 0.1 ns. Single-qubit operations using the optical Stark effect can be performed in about 0.1 ps\cite{Gupta}.
We have also shown that the conditional Faraday rotation can be used to implement a phase gate between qubits protected by error correction codes. In addition, the phase gate can be applied to many pairs of qubits in parallel. Error correction codes and paralled computing make our quantum computing scheme fault-tolerant. The combination of the optical Stark effect\cite{Gupta} with the optical RKKY interaction\cite{Piermarocchi} allows to perform single-qubit operations on Shor's nine-qubit code\cite{Shor1995} in about 0.1 ns.

{\it Acknowledgement}. I thank Ren-Bao Liu, Wang Yao, and Lu J. Sham for useful discussions. This work has been supported by the US NSF DMR-0099572.

\end{document}